\documentclass[11pt]{article}

\usepackage{amssymb}
\usepackage{amsmath}
\usepackage{latexsym}
\usepackage{graphics}
\usepackage{color}

\newcommand{\tr}[1]{\:{\rm tr}\,#1}
\newcommand{\Tr}[1]{\:{\rm Tr}\,#1}

\newcommand{\mbf}[1]{{\boldsymbol {#1} }}
\newcommand{\complex}{{\mathbb C}} 
\newcommand{\zed}{{\mathbb Z}} 
\newcommand{\nat}{{\mathbb N}} 
\newcommand{\real}{{\mathbb R}} 

\def\e{{\,\rm e}\,}

\newcommand{\id}{{1\!\!1}}

\def\ii{{\,{\rm i}\,}}
\def\dd{{\rm d}}

\newcommand{\sect}[1]{\noindent{\bf {#1}.} }
\newcommand{\subsect}[1]{\noindent{\it {#1}.} }
\newcommand{\NO}{\,\mbox{$\circ\atop\circ$}\,} 

\newcommand{\eq}{\begin{equation}}
\newcommand{\eqend}{\end{equation}}
\newcommand{\eqa}{\begin{eqnarray}}
\newcommand{\nonueqa}{\begin{eqnarray*}}
\newcommand{\eqaend}{\end{eqnarray}}
\newcommand{\nonueqaend}{\end{eqnarray*}}

\newcommand{\bma}[1]{\begin{array}{#1}}
\newcommand{\ema}{\end{array}}
\newcommand{\bc}{\begin{center}}
\newcommand{\ec}{\end{center}}

\def\appendix#1{\addtocounter{section}{1}\setcounter{equation}{0}
\renewcommand{\thesection}{\Alph{section}}
\section*{Appendix \thesection\protect\indent \parbox[t]{11.715cm} {#1}}
\addcontentsline{toc}{section}{Appendix \thesection\ \ \ #1} }

\def\Dirac{{D\!\!\!\!/\,}} 
\def\delslash{{\partial\!\!\!/\,}}
\def\pslash{{p\!\!\!/\,}}
\def\kslash{{k\!\!\!/\,}}
\def\qslash{{q\!\!\!/\,}}
\def\Aslash{{A\!\!\!/\,}}

\newif\ifold             \oldtrue

\def\nn{\nonumber}

\def\e{{\,\rm e}\,}

\def\beq{\begin{equation}}
\def\eeq{\end{equation}}
\def\bea{\begin{eqnarray}}
\def\eea{\end{eqnarray}}
\def\bd{\begin{displaymath}}
\def\ed{\end{displaymath}}

\begin{document}

\begin{flushright}
\baselineskip=12pt
HWM--05--08\\
EMPG--05--09\\
hep--th/0505139\\
\hfill{ }\\
May 2005
\end{flushright}

\begin{center}

\baselineskip=20pt

{\Large\bf PERTURBATION THEORY AND TECHNIQUES\footnote{To appear in
    the Quantum Field Theory section of the {\it Encyclopedia of
      Mathematical Physics}, Elsevier, 2006.}}

\baselineskip=12pt

\vspace{0.5 cm}

{\large Richard J.~Szabo}\\

\vspace{0.4 cm}

{\it Department of Mathematics\\ School of Mathematical and Computer
  Sciences\\ Heriot-Watt University\\ Colin Maclaurin Building, Riccarton,
Edinburgh EH14 4AS, U.K.\\{\tt R.J.Szabo@ma.hw.ac.uk}}

\end{center}

\baselineskip=14pt

\bigskip

\sect{1. Introduction}
There are several equivalent formulations of the problem of quantizing
an interacting field theory. The list includes canonical quantization, path
integral (or functional) techniques, stochastic quantization,
``unified'' methods such as the Batalin-Vilkovisky formalism, and
techniques based on the realizations of field theories as low energy
limits of string theory. The problem of obtaining an exact
nonperturbative description of a given quantum field theory is most
often a very difficult one. Perturbative techniques, on the other
hand, are abundant, and common to all of the quantization methods
mentioned above is that they admit particle interpretations in
this formalism.

The basic physical quantities that one wishes to calculate in a
relativistic $(d+1)$-dimensional quantum field theory are the S-matrix
elements
\beq
S_{ba}={}^{~}_{\rm out}\bigl\langle\psi_b(t)\bigm
|\psi_a(t)\bigr\rangle^{~}_{\rm in}
\label{Smatrixgen}\eeq
between in and out states at large positive time $t$. The scattering
operator $\sf S$ is then defined by writing (\ref{Smatrixgen}) in
terms of initial free particle (descriptor) states as
\beq
S_{ba}=:\bigl\langle\psi_b(0)\bigl
|\,{\sf S}\,\bigr|\psi_a(0)\bigr\rangle \ .
\label{Sopdefgen}\eeq
Suppose that the hamiltonian of the given field theory can be written as
$H=H_0+H'$, where $H_0$ is the free part and $H'$ the interaction
hamiltonian. The time evolution of the in and out states are governed
by the total hamiltonian $H$. They can be expressed in terms of
descriptor states which evolve in time with $H_0$ in the interaction
picture and correspond to free particle states. This leads to the
Dyson formula
\beq
{\sf S}={\rm T}~\exp\left(-\ii\int\limits_{-\infty}^\infty\dd t~
H_I(t)\right) \ ,
\label{Dysonformula}\eeq
where T denotes time-ordering and $H_I(t)=\int\dd^d\mbf
x~\mathcal{H}_{\rm int}(\mbf x,t)$ is the interaction hamiltonian in the
interaction picture, with $\mathcal{H}_{\rm int}(\mbf x,t)$ the
interaction hamiltonian density, which deals with essentially free
fields. This formula expresses $\sf S$ in terms of interaction picture
operators acting on free particle states in (\ref{Sopdefgen}) and is
the first step towards Feynman perturbation theory.

For many analytic investigations, such as those which arise in
renormalization theory, one is interested instead in the Green's
functions of the quantum field theory, which measure the response of
the system to an external perturbation. For definiteness, let us
consider a free real scalar field theory in $(d+1)$-dimensions with
lagrangian density
\beq
\mathcal{L}=\mbox{$\frac12$}\,\partial_\mu\phi\,\partial^\mu\phi-
\mbox{$\frac12$}\,m^2\,\phi^2+\mathcal{L}_{\rm int}
\label{scalarLgen}\eeq
where $\mathcal{L}_{\rm int}$ is the interaction lagrangian density which we
assume has no derivative terms. The interaction hamiltonian density is
then given by $\mathcal{H}_{\rm int}=-\mathcal{L}_{\rm
  int}$. Introducing a real scalar source $J(x)$, we define the
normalized ``partition function'' through the vacuum expectation
values
\beq
Z[J]=\frac{\bigl\langle0\bigl|\,{\sf S}[J]\,\bigr|0\bigr
\rangle}{\bigl\langle0\bigl|\,{\sf S}[0]\,\bigr|0\bigr\rangle}
\label{genfndef}\eeq
where $|0\rangle$ is the normalized perturbative vacuum state of the
quantum field theory given by (\ref{scalarLgen}) (defined to be
destroyed by all field annihilation operators), and
\beq
{\sf S}[J]={\rm T}~\exp\left(\ii\int\dd^{d+1}x~\bigl(\mathcal{L}_{\rm int}+
J(x)\,\phi(x)\bigr)\right)
\label{DysonSJ}\eeq
from the Dyson formula. This partition function is the generating
functional for all Green's functions of the quantum field theory, which
are obtained from (\ref{genfndef}) by taking functional derivatives
with respect to the source and then setting $J(x)=0$. Explicitly, in a
formal Taylor series expansion in $J$ one has
\beq
Z[J]=\sum_{n=0}^\infty\frac{\ii^n}{n!}\,\prod_{i=1}^n\,\int
\dd^{d+1}x_i~J(x_i)~G^{(n)}(x_1,\dots,x_n) \ ,
\label{ZJexppos}\eeq
whose coefficients are the Green's functions
\beq
G^{(n)}(x_1,\dots,x_n):=
\frac{\bigl\langle0\bigl|\,{\rm T}\bigl[\exp\left(
\ii\int\dd^{d+1}x~\mathcal{L}_{\rm int}\right)\,\phi(x_1)\cdots
\phi(x_n)\bigr]\,\bigr|0\bigr\rangle}
{\bigl\langle0\bigl|\,{\rm T}~\exp\left(\ii\int
\dd^{d+1}x~\mathcal{L}_{\rm int}\right)\,\bigr|0\bigr\rangle} \ .
\label{Greensfnsdef}\eeq

It is customary to work in momentum space by introducing the Fourier
transforms
\beq
\tilde{J}(k)=\int\dd^{d+1}x~\e^{\ii k\cdot x}~J(x) \ , ~~
\tilde{G}^{(n)}(k_1,\dots,k_n)=\prod_{i=1}^n\,\int\dd^{d+1}x_i~
\e^{\ii k_i\cdot x_i}~G^{(n)}(x_1,\dots,x_n) \ ,
\label{Fourierdefs}\eeq
in terms of which the expansion (\ref{ZJexppos}) reads
\beq
Z[J]=\sum_{n=0}^\infty\frac{\ii^n}{n!}\,\prod_{i=1}^n\,\int
\frac{\dd^{d+1}k_i}{(2\pi)^{d+1}}~\tilde J(-k_i)~
\tilde{G}^{(n)}(k_1,\dots,k_n) \ .
\label{ZJexpmom}\eeq
The generating functional (\ref{ZJexpmom}) can be written as a sum of
Feynman diagrams with source insertions. Diagrammatically, the Green's
function is an infinite series of graphs which can be
represented symbolically as
\begin{equation}
\input{pert1.pstex_t}
\label{GreenFeyngen}\end{equation}
where the $n$ external lines denote the source insertions of momenta
$k_i$ and the bubble denotes the sum over all Feynman diagrams
constructed from the interaction vertices of $\mathcal{L}_{\rm
  int}$.

The Green's functions can also be used to describe scattering
amplitudes, but there are two important differences between the graphs
(\ref{GreenFeyngen}) and those which appear in scattering theory. In
the present case, external lines carry propagators, i.e. the free
field Green's functions
\beq
\Delta(x-y)=\bigl\langle0\bigl|\,{\rm T}\bigl[\phi(x)\,\phi(y)
\bigr]\,\bigr|0\bigr\rangle=\bigl\langle x\bigl|\left(\Box+m^2\right)^{-1}
\bigr|y\bigr\rangle=\int\frac{\dd^{d+1}p}{(2\pi)^{d+1}}~
\frac\ii{p^2-m^2+\ii\epsilon}~\e^{-\ii p\cdot(x-y)}
\label{Deltascalar}\eeq
where $\epsilon\to0^+$ regulates the mass shell
contributions, and their momenta $k_i$ are off-shell in
general ($k_i^2\neq m^2$). By the LSZ theorem, the S-matrix element is
then given by the multiple on-shell residue of the Green's function in
momentum space as
\bea
\left\langle\left.\left. k_1',\dots,k_n'\right|\,{\sf S}-\id\,
\right|k^{~}_1,\dots,k^{~}_l\right\rangle&=&
\lim_{\stackrel{\scriptstyle k_1',\dots,k_n'\to m^2}
{\scriptstyle k^{~}_1,\dots,k^{~}_l\to
  m^2}}~\prod_{i=1}^n\mbox{$\frac1{\ii\sqrt{c_i'}}$}\,
\left(k_i^{\prime\,2}-m^2\right)\,\prod_{j=1}^l\mbox{$\frac1{\ii
\sqrt{c^{~}_j}}$}\,
\left(k_j^2-m^2\right)\nonumber\\ && \times\,
\tilde G^{(n+m)}\left(-k_1',\dots,-k_n',k^{~}_1,\dots,k^{~}_l
\right) \ ,
\label{SGreensrelgen}\eea
where $\ii c_i',\ii c_j^{~}$ are the residues of the corresponding
particle poles in the exact two-point Green's function.

This article deals with the formal development and computation of
perturbative scattering amplitudes in relativistic quantum field
theory, along the lines outlined above. Initially we deal only
with real scalar field theories of the sort (\ref{scalarLgen}) in
order to illustrate the concepts and technical tools in as simple and
concise a fashion as possible. These techniques are common to most
quantum field theories. Fermions and gauge theories are then
separately treated afterwards, focusing on the methods which are
particular to them.

\bigskip

\sect{2. Diagrammatics}
The pinnacle of perturbation theory is the technique of Feynman
diagrams. Here we develop the basic machinery in a quite general
setting and use it to analyse some generic features of the terms
comprising the perturbation series.

\smallskip

\subsect{Wick's theorem}
The Green's functions (\ref{Greensfnsdef}) are defined in terms of
vacuum expectation values of time-ordered products of the scalar field
$\phi(x)$ at different spacetime points. Wick's theorem expresses such
products in terms of normal-ordered products, defined by placing each
field creation operator to the right of each field annihilation
operator, and in terms of two-point Green's functions
(\ref{Deltascalar}) of the free field theory (propagators). The
consequence of this theorem is the haffnian formula
\beq
\bigl\langle0\bigl|\,{\rm T}\bigl[\phi(x_1)\cdots\phi(x_n)\bigr]\,
\bigr|0\bigr\rangle~=~\left\{\begin{matrix}0&,&n=2k-1 \ , \\
\,\sum\limits_{\pi\in S_{2k}}~\prod\limits_{i=1}^k\,
\bigl\langle0\bigl|\,{\rm T}\bigl[\phi(x_{\pi(2i-1)})\,
\phi(x_{\pi(2i)})\bigr]\,\bigr|0\bigr\rangle&,&n=2k \ . \end{matrix}
\right.
\label{Wicksthm}\eeq

The formal Taylor series expansion of the scattering operator $\sf S$
may now be succinctly summarized into a diagrammatic notation by using
Wick's theorem. For each spacetime integration $\int\dd^{d+1}x_i$ we
introduce a vertex with label $i$, and from each vertex there emanates
some lines corresponding to field insertions at the point $x_i$. If
the operators represented by two lines appear in a two-point function
according to (\ref{Wicksthm}), i.e. they are {\it contracted}, then
these two lines are connected together. The $\sf S$ operator is then
represented as a sum over all such Wick diagrams,
bearing in mind that topologically equivalent diagrams correspond to
the same term in $\sf S$. Two diagrams are said to have the same {\it
  pattern} if they differ only by a permutation of their vertices. For
any diagram $\mathfrak{D}$ with $n(\mathfrak{D})$ vertices, the number
of ways of interchanging vertices in $n(\mathfrak{D})!$. The number of
diagrams per pattern is always less than this number. The {\it
  symmetry number} $S(\mathfrak{D})$ of $\mathfrak{D}$ is the number
of permutations of vertices that give the same diagram. The number of
diagrams with the pattern of $\mathfrak{D}$ is then
$n(\mathfrak{D})!/S(\mathfrak{D})$.

In a given pattern, we write the contribution to $\sf S$ of a single
diagram $\mathfrak{D}$ as
$\frac1{n(\mathfrak{D})!}\,\NO\theta(\mathfrak{D})\NO$, where the
combinatorial factor comes from the Taylor expansion of $\sf S$,
the colons denote normal ordering of quantum operators, and
$\NO\theta(\mathfrak{D})\NO$ contains spacetime integrals over
normal-ordered products of the fields. Then
all diagrams with the pattern of $\mathfrak{D}$ contribute
$\NO\theta(\mathfrak{D})\NO/S(\mathfrak{D})$ to $\sf S$. Only the {\it
  connected} diagrams $\mathfrak{D}_r$,
$r\in\nat$ (those in which every vertex is connected to every other
vertex) contribute and we can write the scattering operator in a
simple form which eliminates contributions from all disconnected
diagrams as
\beq
{\sf S}=\NO\exp\left(\,\sum_{r=1}^\infty\frac{\theta(\mathfrak{D}_r)}
{S(\mathfrak{D}_r)}\right)\NO \ .
\label{Sdiagconn}\eeq

\smallskip

\subsect{Feynman rules}
Feynman diagrams in momentum space are defined from the Wick diagrams
above by dropping the labels on vertices (and also the symmetry
factors $S(\mathfrak{D})^{-1}$), and
by labelling the external lines by the momenta of the initial and final
particles that the corresponding field operators annihilate. In a spacetime
interpretation, external lines represent on-shell physical particles
while internal lines of the graph represent off-shell virtual
particles ($k^2\neq m^2$). Physical particles interact via the
exchange of virtual particles. An arbitrary diagram is then calculated
via the Feynman rules
\beq
\!\!\!\!\!\!\!\!\!\!\!\!\!\!\!\!\!\!\!\!\!\!\!\!\!\!\!
\!\!\!\!\!\!\!\!\!\!\!\!\!\!\!\!\!\!\!\!\!\!\!\!\!\!\!
\!\!\!\!\!\!\!\!\!\!\!\!\!\!\!\!\!\!\!\!\!\!\!\!\!\!\!\input{pert2.pstex_t}
\label{Feynrulesscalarprop}\eeq
for a monomial interaction $\mathcal{L}_{\rm int}=\frac
g{n!}\,\phi^n$.

\smallskip

\subsect{Irreducible Green's functions}
A {\it one-particle irreducible (1PI)} or {\it proper} Green's
function is given by a sum of diagrams in which each diagram cannot be
separated by cutting one internal line. In momentum space, it is
defined without the overall momentum conservation delta-function
factors and without propagators on external lines. For example, the
two particle 1PI Green's function
\beq
\input{pert3.pstex_t}
\label{1PI2part}\eeq
is called the {\it self-energy}. If $G(k)$ is the complete two-point
function in momentum space, then one has
\beq
\input{pert4.pstex_t}
\label{2ptfullsum}\eeq
and thus it suffices to calculate only 1PI diagrams.

The 1PI effective action, defined by the Legendre transformation
$\Gamma[\phi]:=-\ii\ln Z[J]-\int\dd^{d+1}x~J(x)\,\phi(x)$ of
(\ref{genfndef}), is the generating functional
for proper vertex functions and it can be represented as a functional
of only the vacuum expectation value of the field $\phi$, i.e. its classical
value. In the semi-classical (WKB) approximation, the one-loop
effective action is given by
\bea
\Gamma[\phi]&=&S[\phi]+\mbox{$\frac{\ii\hbar}2$}\,\Tr\ln\bigl(
1+\Delta\,V^{\prime\prime}[\phi]\bigr)+O\left(\hbar^2\right)\nn\\
&=&S[\phi]+\ii\hbar\,\sum_{n=1}^\infty\frac{(-1)^n}{2n}\,
\prod_{i=1}^n\,\int\dd^{d+1}x_i~\Delta(x_i-x_{i+1})\,
V^{\prime\prime}\bigl[\phi(x_{i+1})\bigr]+O\left(\hbar^2\right)
\label{1PIeffaction}\eea
where we have denoted $S[\phi]=\int\dd^{d+1}x~\mathcal{L}$ and
$V[\phi]=-\mathcal{L}_{\rm int}$, and for each term in the infinite
series we define $x_{n+1}:=x_1$. The first term in
(\ref{1PIeffaction}) is the classical contribution and it can be
represented in terms of connected tree diagrams. The second term is
the sum of contributions of
one-loop diagrams constructed from $n$ propagators $-\ii\Delta(x-y)$
and $n$ vertices $-\ii V^{\prime\prime}[\phi]$. The expansion may
be carried out to all orders in terms of connected Feynman diagrams,
and the result of the above Legendre transformation is to select only
the one-particle irreducible diagrams and to replace the classical
value of $\phi$ by an arbitrary argument. All information about the
quantum field theory is encoded in this effective action.

\smallskip

\subsect{Parametric representation}
Consider an arbitrary proper Feynman diagram $\mathfrak{D}$ with $n$
internal lines and $v$ vertices. The number $\ell$ of independent
loops in the diagram is the number of independent internal momenta in
$\mathfrak{D}$ when conservation laws at each vertex have been taken
into account, and it is given by $\ell=n+1-v$. There is an independent
momentum integration variable $k_i$ for each loop, and a propagator
for each internal line as
in (\ref{Feynrulesscalarprop}). The contribution of $\mathfrak{D}$ to
a proper Green's function with $r$ incoming external momenta $p_i$,
with $\sum_{i=1}^rp_i=0$, is given by
\beq
\tilde I_{\mathfrak D}(p)=\frac{V(\mathfrak{D})}
{S(\mathfrak{D})}\,\prod_{i=1}^n\,\int\frac{\dd^{d+1}k_i}{(2\pi)^{d+1}}~
\frac\ii{k_i^2-m^2+\ii\epsilon}~\prod_{j=1}^v\,(2\pi)^{d+1}\,
\delta^{(d+1)}\left(P_j-K_j\right)
\label{Igencontr}\eeq
where $V(\mathfrak{D})$ contains all contributions from the
interaction vertices of $\mathcal{L}_{\rm int}$, and $P_j$
(resp. $K_j$) is the sum of incoming external momenta
$p_{l_j}$ (resp. internal momenta $k_{l_j}$) at vertex $j$ with
respect to a fixed chosen orientation of the lines of the graph. After
resolving the delta-functions in terms of independent internal
loop momenta $k_1,\dots,k_\ell$ and dropping the overall momentum
conservation delta-function along with the symmetry and vertex factors
in (\ref{Igencontr}), one is left with a set of momentum space
integrals
\beq
I_{\mathfrak D}(p)=\prod_{i=1}^\ell\,\int
\frac{\dd^{d+1} k_i}{(2\pi)^{d+1}}~\prod_{j=1}^n
\frac\ii{a_j(k,p)+\ii\epsilon}
\label{Igenafterdelta}\eeq
where $a_j(k,p)$ are functions of both the internal and external
momenta.

It is convenient to exponentiate propagators using the Schwinger
parametrization
\beq
\frac\ii{a_j+\ii\epsilon}=\int\limits_0^\infty\dd\alpha_j~
\e^{\ii\alpha_j(a_j+\ii\epsilon)} \ ,
\label{Schwingerpars}\eeq
and after some straightforward manipulations one may write the Feynman
parametric formula
\beq
\prod_{j=1}^n\frac\ii{a_j(k,p)+\ii\epsilon}=(n-1)!\,\prod_{j=1}^n\,
\int\limits_0^1\dd\alpha_j~\frac{\delta\left(1-\sum_j\alpha_j\right)}
{D_{\mathfrak{D}}(k;\alpha,p)^n}
\label{Feynpars}\eeq
where
$D_{\mathfrak{D}}(k;\alpha,p):=\sum_j\alpha_j[a_j(k,p)+\ii\epsilon]$
is generically a quadratic form
\beq
D_{\mathfrak{D}}(k;\alpha,p)=\frac12\,\sum_{i,j=1}^\ell k_i\cdot
Q_{ij}(\alpha)k_j+\sum_{i=1}^\ell L_i(p)\cdot k_i+\lambda\left(p^2
\right) \ .
\label{Dquadform}\eeq
The positive symmetric matrix $Q_{ij}$ is independent of the external
momenta $p_l$, invertible, and has non-zero eigenvalues
$Q_1,\dots,Q_\ell$. The vectors $L_i^\mu$ are linear combinations of
the $p_j^\mu$, while $\lambda(p^2)$ is a function of only the Lorentz
invariants $p_i^2$. After some further elementary manipulations, the
loop diagram contribution (\ref{Igenafterdelta}) may be written as
\bea
I_{\mathfrak D}(p)&=&(n-1)!\,
\prod_{j=1}^n\,\int\limits_0^1\dd\alpha_j~
\prod_{i=1}^\ell\frac1{Q_i(\alpha)^2}\,\int\frac{\dd^{d+1}k_i}{(2\pi)^{d+1}}~
\delta\left(1-\mbox{$\sum_j$}\,\alpha_j\right)\nn\\ &&
\times\,\left(\mbox{$\frac12\,\sum_i$}\,k_i^2+\lambda\left(p^2
\right)-\mbox{$\frac12\,\sum_{i,j}$}\,L_i(p)\cdot Q^{-1}(\alpha)_{ij}L_j(p)
\right)^{-n} \ .
\label{Iaftermans}\eea
Finally, the integrals over the loop momenta $k_i$ may be performed by
Wick rotating them to euclidean space and using the fact that the
combination of $\ell$ integrations in $\real^{d+1}$ has
$O((d+1)\ell)$ rotational invariance. The contribution from the entire
Feynman diagram $\mathfrak D$ thereby reduces to the calculation of
the parametric integrals
\beq
I_{\mathfrak D}(p)=\frac{\Gamma\left(n-\frac{(d+1)\ell}2\right)}
{(2\pi)^{\frac{(d+1)\ell}2}\ii^{d\,\ell}}
\,\prod_{j=1}^n\,\int\limits_0^1\dd\alpha_j~
\prod_{i=1}^\ell\frac1{Q_i(\alpha)^2}~\frac{\delta\left(
1-\sum_j\alpha_j\right)}{\left(\lambda\left(p^2
\right)-\frac12\,\sum_{i,j}L_i(p)\cdot Q^{-1}(\alpha)_{ij}L_j(p)
\right)^{n-\frac{(d+1)\ell}2}}
\label{Igenfinal}\eeq
where $\Gamma(s)$ is the Euler gamma-function.

\bigskip

\sect{3. Regularization}
The parametric representation (\ref{Igenfinal}) is generically
convergent when $2n-(d+1)\ell>0$. When divergent, the infinities arise
from the lower limits of integration $\alpha_j\to0$. This is just the
parametric representation of the large $k$ divergence of the original
Feynman amplitude (\ref{Igencontr}). Such ultraviolet divergences
plague the very meaning of a quantum field theory and must be dealt
with in some way. We will now quickly tour the standard methods of ultraviolet
regularization for such loop integrals, which is prelude to the
renormalization program that removes the divergences (in a renormalizable
field theory). Here we consider regularization simply as a means of
justification for the various formal manipulations that are used in
arriving at expressions such as (\ref{Igenfinal}).

\smallskip

\subsect{Momentum cutoff}
Cutoff regularization introduces a mass scale $\Lambda$ into the
quantum field theory and throws away the Fourier modes of the fields
for spatial momenta $\mbf k$ with $|\mbf k|>\Lambda$. This regularization
spoils Lorentz invariance. It is also non-local. For example, if we
restrict to a hypercube in momentum space, so that $|k_i|<\Lambda$ for
$i=1,\dots,d$, then $\int_{|\mbf k|<\Lambda}\,\frac{\dd^d\mbf
  k}{(2\pi)^d}~\e^{\ii\mbf k\cdot\mbf
  x}=\prod_{i=1}^d\,\sin(\Lambda\,x^i)/\pi\,x^i$ which is a delta-function
in the limit $\Lambda\to\infty$ but is non-local for
$\Lambda<\infty$. The regularized field theory is finite order by
order in perturbation theory and depends on the cutoff~$\Lambda$.

\smallskip

\subsect{Lattice regularization}
We can replace the spatial continuum by a lattice $\mathfrak{L}$ of
rank $d$ and define a lagrangian on $\mathfrak{L}$ by
\beq
{L}_{\mathfrak{L}}=\frac12\,\sum_{i\in{\rm S}(\mathfrak{L})}
\dot\phi_i^{\,2}+J\,\sum_{\langle i,j\rangle\in{\rm L}(\mathfrak{L})}
\phi_i\,\phi_j+\sum_{i\in{\rm S}(\mathfrak{L})}V(\phi_i) \ ,
\label{latticeL}\eeq
where ${\rm S}(\mathfrak{L})$ is the set of sites $i$ of the lattice on
each of which is situated a time-dependent function $\phi_i$, and
${\rm L}(\mathfrak{L})$ is the collection of links connecting
pairs $\langle i,j\rangle$ of nearest-neighbour sites $i,j$ on
$\mathfrak{L}$. The regularized field theory is now local, but still
has broken Lorentz invariance. In particular, it suffers from broken
rotational symmetry. If $\mathfrak{L}$ is hypercubic with lattice
spacing $a$, i.e. $\mathfrak{L}=(\zed\,a)^d$, then the momentum cutoff
is at $\Lambda=a^{-1}$.

\smallskip

\subsect{Pauli-Villars regularization}
We can replace the propagator $\ii(k^2-m^2+\ii\epsilon)^{-1}$ by
$\ii(k^2-m^2+\ii\epsilon)^{-1}+\ii\sum_{j=1}^Nc^{~}_j(k^2-M_j^2+\ii\epsilon)^{-1}$,
where the masses $M_j\gg m$ are identified with the momentum cutoff
as $\min\{M_j\}=\Lambda\to\infty$. The mass-dependent coefficients $c^{~}_j$
are
chosen to make the modified propagator decay rapidly as $(k^2)^{-N-1}$
at $k\to\infty$, which gives the $N$ equations
$(m^2)^i+\sum_j\,c^{~}_j(M_j^2)^i=0$, $i=0,1,\dots,N-1$. This
regularization preserves Lorentz invariance (and other symmetries that
the field theory may possess) and is local in the following sense. The
modified propagator can be thought of as arising through the
alteration of the lagrangian density (\ref{scalarLgen}) by $N$
additional scalar fields $\varphi_j$ of masses $M_j$ with
\beq
\mathcal{L}_{\rm PV}=\mbox{$\frac12$}\,\partial_\mu\phi\,\partial^\mu\phi-
\mbox{$\frac12$}\,m^2\,\phi^2+\sum_{j=1}^N\left(\mbox{$\frac12$}\,
\partial_\mu\varphi_j\,\partial^\mu\varphi_j-
\mbox{$\frac12$}\,M_j^2\,\varphi_j^2\right)+\mathcal{L}_{\rm
int}[\Phi]
\label{PVLgen}\eeq
where $\Phi:=\phi+\sum_j\,\sqrt{c_j}~\varphi_j$. The contraction of
the $\Phi$ field thus produces the required propagator. However, the
$c_j$'s as computed above are generically negative numbers and so the
lagrangian
density (\ref{PVLgen}) is not hermitian (as $\Phi\neq\Phi^\dag$). It is
possible to make (\ref{PVLgen}) {\it formally} hermitian by redefining
the inner product on the Hilbert space of physical states, but this
produces negative norm states. This is no problem at energy scales
$E\ll M_j$ on which the extra particles decouple and the negative
probability states are invisible.

\smallskip

\subsect{Dimensional regularization}
Consider a euclidean space integral $\int\dd^4k~(k^2+a^2)^{-r}$ arising
after Wick rotation from some loop diagram in $(3+1)$-dimensional
scalar field theory. We replace this integral by its $D$-dimensional
version
\beq
\int\frac{\dd^Dk}{\left(k^2+a^2\right)^r}=\frac{\pi^{\frac D2}\,
\left(a^2\right)^{\frac D2-r}}{(r-1)!}~\Gamma\left(r-
\mbox{$\frac D2$}\right) \ .
\label{intDreplace}\eeq
This integral is absolutely convergent for $D<2r$. We can analytically
continue the result of this integration to the complex plane
$D\in\complex$. As an analytic function, the only singularities of the
Euler function $\Gamma(z)$ are poles at $z=0,-1,-2,\dots$. In
particular, $\Gamma(z)$ has a simple pole at $z=0$ of residue~$1$. If
we write $D=4+\epsilon$ with $|\epsilon|\to0$, then the integral
(\ref{intDreplace}) is proportional to $\Gamma(r-2-\frac\epsilon2)$
and $\epsilon$ plays the role of the regulator here. This
regularization is Lorentz invariant (in $D$ dimensions) and is
distinguished as having a dimensionless regularization parameter
$\epsilon$. This parameter is related to the momentum cutoff $\Lambda$
by $\epsilon^{-1}=\ln(\Lambda/m)$, so that the limit $\epsilon\to0$
corresponds to $\Lambda\to\infty$.

\smallskip

\subsect{Infrared divergences}
Thus far we have only considered the ultraviolet behaviour of loop
amplitudes in quantum field theory. When dealing with massless
particles ($m=0$ in (\ref{scalarLgen})) one has to further worry about
divergences arising from the $k\to0$ regions of Feynman
integrals. After Wick rotation to euclidean momenta, one can show that
no singularities arise in a given Feynman diagram as some of its
internal masses vanish provided that all vertices have superficial
degree of divergence~$d+1$, the external momenta are not exceptional
(i.e. no partial sum of the incoming momenta $p_i$ vanishes), and
there is at most one soft external momentum. This result assumes that
renormalization has been carried out at some fixed euclidean
point. The extension of this property when the external momenta are
continued to physical on-shell values is difficult. The
Kinoshita-Lee-Nauenberg theorem states that, as a consequence of
unitarity, transition probabilities in a theory involving massless
particles are finite when the sum over all degenerate states (initial
and final) is taken. This is true order by order in perturbation
theory in bare quantities or if minimal subtraction renormalization is
used (to avoid infrared or mass singularities in the renormalization
constants).

\bigskip

\sect{4. Fermion Fields}
We will now leave the generalities of our pure scalar field theory and
start considering the extensions of our previous considerations to
other types of particles. Henceforth we will primarily deal with the
case of $(3+1)$-dimensional spacetime. We begin by indicating how the
rudiments of perturbation theory above apply to the case of Dirac
fermion fields. The lagrangian density is
\beq
\mathcal{L}_F=\overline{\psi}\,(\ii\delslash-m)\psi+\mathcal{L}'
\label{fermaction}\eeq
where $\psi$ are four-component Dirac fermion fields in
$(3+1)$-dimensions, $\overline{\psi}:=\psi^\dag\gamma^0$ and
$\delslash=\gamma^\mu\,\partial_\mu$ with $\gamma^\mu$ the generators
of the Clifford algebra
$\{\gamma^\mu,\gamma^\nu\}=2\eta^{\mu\nu}$. The lagrangian density
$\mathcal{L}'$ contains couplings of the Dirac fields to other field
theories, such as the scalar field theories considered previously.

Wick's theorem for anticommuting Fermi fields leads to the pfaffian
formula
\beq
\bigl\langle0\bigl|\,{\rm T}\bigl[\psi(1)\cdots\psi(n)\bigr]\,
\bigr|0\bigr\rangle~=~\left\{\begin{array}{l}0~~,~~n=2k-1 \ , \\
\frac1{2^k\,k!}\,\sum\limits_{\pi\in S_{2k}}{\rm sgn}(\pi)
\,\prod\limits_{i=1}^k\,
\bigl\langle0\bigl|\,{\rm T}\bigl[\psi\bigl(\pi(2i-1)\bigr)\,
\psi\bigl(\pi(2i)\bigr)\bigr]\,\bigr|0\bigr\rangle~~,~~n=2k \end{array}
\right.
\label{Wicksthmferm}\eeq
where for compactness we have written in the argument of $\psi(i)$ the
spacetime coordinate, the Dirac index, and a discrete index which
distinguishes $\psi$ from $\overline{\psi}$. The non-vanishing
contractions in (\ref{Wicksthmferm}) are determined by the free
fermion propagator
\beq
\Delta_F(x-y)=\bigl\langle0\bigl|\,{\rm T}\bigl[\psi(x)\,
\overline{\psi}(y)\bigr]\,\bigr|0\bigr\rangle=\bigl\langle x
\bigl|(\ii\delslash-m)^{-1}\bigr|y\bigr\rangle=\ii\int
\frac{\dd^4p}{(2\pi)^4}~\frac{\pslash+m}{p^2-m^2+\ii\epsilon}~
\e^{-\ii p\cdot(x-y)} \ .
\label{freefermprop}\eeq
Perturbation theory now proceeds exactly as before. Suppose that the
coupling lagrangian density in (\ref{fermaction}) is of the form
$\mathcal{L}'=\overline{\psi}(x)\,V(x)\,\psi(x)$. Both the Dyson formula
(\ref{Dysonformula}) and the diagrammatic formula (\ref{Sdiagconn})
are formally the same in this instance. For example, in the formal expansion
in powers of $\int\dd^4x~\mathcal{L}'$, the vacuum-to-vacuum
amplitude (the denominator in (\ref{genfndef})) will contain field
products of the form $\prod_{i=1}^n\,\int\dd^4x_i~\langle0|\,{\rm
  T}[\,\overline{\psi}(x_i)\,V(x_i)\,\psi(x_i)]\,|0\rangle$ which
correspond to fermion loops. Before applying Wick's theorem the fields
must be rearranged as
$\tr\prod_{i=1}^n\,V(x_i)\,\psi(x_i)\,\overline{\psi}(x_{i+1})$ (with
$x_{n+1}:=x_1$), where $\tr$ is the $4\times4$ trace over spinor
indices. This reordering introduces the familiar minus sign for a
closed fermion loop, and one has
\beq
\!\!\!\!\!\!\!\!\!\!\!\!\!\!\!\!\!\!
\!\!\!\!\!\!\!\!\!\!\!\!\!\!\!\!\!\!
\!\!\!\!\!\!\!\!\!\!\!\!\!\!\!\!\!\!
\!\!\!\!\!\!\!\!\!\!\!\!\!\!\!\!\!\!
\!\!\!\!\!\!\!\!\!\!\!\!\!\!\!\!\!\!
\!\!\!\!\!\!\!\!\!\!\!\!\!\!\!\!\!\!
\!\!\!\!\!\!\!\!\!\!\!\!\!\!\!\!\!\!
\!\!\!\!\!\!\!\!\!\!\!\!\!\!\!\!\!\!
\!\!\!\!\!\!\!\!\!\!\!\!\!\!\!\!\!\!
\!\!\!\!\!\!\!\!\!\!\!\!\!\!\!\!\!\!
\input{pert5.pstex_t}
\label{fermloopminus}\eeq

Feynman rules are now described as follows. Fermion lines are oriented
to distinguish a particle from its corresponding antiparticle, and
carry both a four-momentum label $p$ as well as a spin polarization index
$r=1,2$. Incoming fermions (resp. antifermions) are described by
the wavefunctions $u_{p}^{(r)}$
(resp. $\overline{v}{}^{\,(r)}_{p}$), while outgoing fermions
(resp. antifermions) are described by the wavefunctions
$\overline{u}{}_{p}^{\,(r)}$ (resp. $v^{(r)}_{p}$). Here
$u_{p}^{(r)}$ and $v^{(r)}_{p}$ are the
classical spinors, i.e. the positive and negative energy solutions of
the Dirac equation
$(\pslash-m)u^{(r)}_{p}=(\pslash+m)v^{(r)}_{p}=0$. Matrices are
multiplied along a Fermi line, with the head of the arrow on the
left. Closed fermion loops produce an overall minus sign as in
(\ref{fermloopminus}), and the multiplication rule gives the trace of
Dirac matrices along the lines of the loop.
Unpolarized scattering amplitudes are summed
over the spins of final particles and averaged over the spins of
initial particles using the completeness relations for spinors
\beq
\sum_{r=1,2}u_p^{(r)}\,\overline{u}{}_p^{\,(r)}=\pslash+m  \ , ~~
\sum_{r=1,2}v_p^{(r)}\,\overline{v}{}_p^{\,(r)}=\pslash-m \ ,
\label{spinorcomplrel}\eeq
leading to basis independent results. Polarized amplitudes are
computed using the spinor bilinears
$\overline{u}{}_p^{\,(r)}\,\gamma^\mu
u_p^{(s)}=\overline{v}{}_p^{\,(r)}\,\gamma^\mu
v_p^{(s)}=2p^\mu\,\delta^{rs}$, $\overline{u}{}_p^{\,(r)}\,
u_p^{(s)}=-\overline{v}{}_p^{\,(r)}\,
v_p^{(s)}=2m\,\delta^{rs}$, and $\overline{u}{}_p^{\,(r)}\,
v_p^{(s)}=0$.

When calculating fermion loop integrals using dimensional
regularization, one utilizes the Dirac algebra in $D$ dimensions
\bea
\gamma^\mu\,\gamma_\mu~=~\eta^\mu_{~\mu}&=&D \ , \nn\\
\gamma_\mu\,\pslash\,\gamma^\mu&=&(2-D)\,\pslash \ , \nn\\
\gamma^\mu\,\pslash\,\kslash\,\gamma_\mu&=&4p\cdot k+(D-4)\,
\pslash\,\kslash \ , \nn\\\gamma^\mu\,\pslash\,\kslash\,\qslash\,
\gamma_\mu&=&-2\qslash\,\kslash\,\pslash-(D-4)\,\pslash\,\kslash\,
\qslash \ , \nn\\ \tr\id~=~4 \ , ~~ \tr\gamma^{\mu_1}
\cdots\gamma^{\mu_{2k-1}}&=&0 \ , ~~ \tr\gamma^\mu\,
\gamma^\nu~=~4\eta^{\mu\nu} \ , \nn\\\tr\gamma^\mu\,
\gamma^\nu\,\gamma^\rho\,\gamma^\sigma&=&4\left(\eta^{\mu\nu}\,
\eta^{\rho\sigma}-\eta^{\mu\rho}\,\eta^{\nu\sigma}+\eta^{\mu\sigma}\,
\eta^{\nu\rho}\right) \ .
\label{DdimDiracalg}\eea
Specific to $D=4$ dimensions are the trace identities
\beq
\tr\gamma^5=\tr\gamma^\mu\,\gamma^\nu\,\gamma^5=0\ , ~~
\tr\gamma^\mu\,\gamma^\nu\,\gamma^\rho\,\gamma^\sigma\,\gamma^5=
-4\ii\epsilon^{\mu\nu\rho\sigma}
\label{4Dtraceids}\eeq
where $\gamma^5:=\ii\gamma^0\,\gamma^1\,\gamma^2\,\gamma^3$. Finally,
loop diagrams evaluated with the fermion propagator
(\ref{freefermprop}) require a generalization of the momentum space
integral (\ref{intDreplace}) given by
\beq
\int\frac{\dd^Dk}{(2\pi)^D}~\frac1{\left(k^2+2k\cdot p+a^2+\ii\epsilon
\right)^r}=\frac{\ii(-\pi)^{\frac D2}\,\Gamma\left(r-\frac D2\right)}
{(2\pi)^D\,(r-1)!}~\frac1{\left(a^2-p^2+\ii\epsilon
\right)^{r-\frac D2}} \ .
\label{fermintDgen}\eeq
{}From this formula we can extract expressions for more complicated Feynman
integrals which are tensorial, i.e. which contain products of momentum
components $k^\mu$ in the numerators of their integrands, by differentiating
(\ref{fermintDgen}) with respect to the external momentum $p^\mu$.

\bigskip

\sect{5. Gauge Fields}
The issues we have dealt with thus far have interesting difficulties
when dealing with gauge fields. We will now discuss some general
aspects of the perturbation expansion of gauge theories using as
prototypical examples quantum electrodynamics (QED) and quantum
chromodynamics (QCD) in four spacetime dimensions.

\smallskip

\subsect{Quantum electrodynamics}
Consider the QED lagrangian density
\beq
\mathcal{L}_{\rm QED}=-\mbox{$\frac14$}\,F_{\mu\nu}\,F^{\mu\nu}+
\mbox{$\frac12$}\,\mu^2\,A_\mu\,A^\mu+\overline{\psi}\,(\ii
\delslash-e\,\Aslash-m)\psi
\label{LQEDdef}\eeq
where $A_\mu$ is a $U(1)$ gauge field in $(3+1)$-dimensions and
$F_{\mu\nu}=\partial_\mu A_\nu-\partial_\nu A_\mu$ is its field
strength tensor. We have added a small mass term $\mu\to0$ for the
gauge field which at the end of calculations should be taken to vanish
in order to describe real photons (as opposed to the {\it soft}
photons described by (\ref{LQEDdef})). This is done in order to cure the
infrared divergences generated in scattering amplitudes due to the
masslessness of the photon, i.e. the long range nature of the
electromagnetic interaction. The Bloch-Nordsieck theorem in QED states
that infrared divergences cancel for physical processes, i.e. for
processes with an arbitrary number of undetectable soft photons.

Perturbation theory proceeds in the usual way via the Dyson formula,
Wick's theorem, and Feynman diagrams. The gauge field propagator is
given by
\beq
\bigl\langle0\bigl|\,{\rm T}\bigl[A_\mu(x)\,A_\nu(y)\bigr]\,
\bigr|0\bigr\rangle=\bigl\langle x\bigl|\bigl[\eta_{\mu\nu}\,
\left(\Box+\mu^2\right)-\partial_\mu\,\partial_\nu\bigr]^{-1}\bigr|y\bigr
\rangle=\ii\int\frac{\dd^4p}{(2\pi)^4}~\frac{-\eta_{\mu\nu}+
\frac{p_\mu\,p_\nu}{\mu^2}}{p^2-\mu^2+\ii\epsilon}~
\e^{-\ii p\cdot(x-y)}
\label{QEDAprop}\eeq
and is represented by a wavy line. The fermion-fermion-photon vertex
is
\beq
\!\!\!\!\!\!\!\!\!\!\!\!\!\!\!\!\!\!
\!\!\!\!\!\!\!\!\!\!\!\!\!\!\!\!\!\!
\input{pert6.pstex_t}
\label{ffpvertex}\eeq
An incoming (resp. outgoing) soft photon of momentum $k$ and
polarization $r$ is described by the wavefunction $e_\mu^{(r)}(k)$
(resp. $e_\mu^{(r)}(k)^*$), where the polarization vectors
$e_\mu^{(r)}(k)$, $r=1,2,3$ solve the vector field wave equation
$(\Box+\mu^2)A_\mu=\partial_\mu A^\mu=0$ and obey the orthonormality
and completeness conditions
\beq
e^{(r)}(k)^*\cdot e^{(s)}(k)=-\delta^{rs} \ , ~~\sum_{r=1}^3
e_\mu^{(r)}(k)\,e_\nu^{(r)}(k)^*=-\eta_{\mu\nu}+\frac{k_\mu\,
k_\nu}{\mu^2} \ ,
\label{photonpolconds}\eeq
along with $k\cdot e^{(r)}(k)=0$. All vector indices are contracted
along the lines of the Feynman graph. All other Feynman rules are as
previously.

\smallskip

\subsect{Quantum chromodynamics}
Consider nonabelian gauge theory in $(3+1)$-dimensions minimally
coupled to a set of fermion fields $\psi^A$, $A=1,\dots,N_f$ each
transforming in the fundamental representation of the gauge group $G$
whose generators $T^a$ satisfy the commutation relations
$[T^a,T^b]=f^{abc}\,T^c$. The lagrangian density is given by
\beq
\mathcal{L}_{\rm QCD}=-\mbox{$\frac14$}\,F_{\mu\nu}^a\,F^{a\,\mu\nu}+
\mbox{$\frac1{2\alpha}$}\,\left(\partial_\mu
  A^{a\,\mu}\right)^2+\partial_\mu\overline{\eta}\,D^\mu\eta+
\sum_{A=1}^{N_f}\,
\overline{\psi}{}^{\,A}\,(\ii\Dirac-m_A)\psi^A \ ,
\label{LQCDdef}\eeq
where $F_{\mu\nu}^a=\partial^{~}_\mu A_\nu^a-\partial^{~}_\nu
A_\mu^a+f^{abc}\,A_\mu^b\,A_\nu^c$ and $D_\mu=\partial_\mu+\ii
e\,R(T^a)\,A_\mu^a$ with $R$ the pertinent representation of $G$
($R(T^a)_{bc}=f^a_{~bc}$ for the adjoint representation and
$R(T^a)=T^a$ for the fundamental representation). The
first term is the Yang-Mills lagrangian density, the second term is
the covariant gauge-fixing term, and the third term contains the
Faddeev-Popov ghost fields $\eta$ which transform in the adjoint
representation of the gauge group.

Feynman rules are straightforward to write down and are given by
\beq
\!\!\!\!\!\!\!\!\!\!\!\!\!\!\!\!\!\!
\!\!\!\!\!\!\!\!\!\!\!\!\!\!\!\!\!\!
\!\!\!\!\!\!\!\!\!\!\!\!\!\!\!\!\!\!
\!\!\!\!\!\!\!\!\!\!\!\!\!\!\!\!\!\!
\input{pert7.pstex_t}
\label{QCDFeynrules}\eeq
where wavy lines represent gluons and dashed lines represent
ghosts. Feynman rules for the fermions are exactly as before, except
that now the vertex (\ref{ffpvertex}) is multiplied by the colour
matrix $T^a$. All colour indices are contracted along the lines of the
Feynman graph. Colour factors may be simplified by using the identities
\beq
\qquad \Tr R^a\,R^b=\mbox{$\frac{\dim R}{\dim G}$}\,C_2(R)~\delta^{ab} \ ,
{}~~ R^a\,R^a=C_2(R) \ , ~~ R^a\,R^b\,R^a=\left(C_2(R)-
\mbox{$\frac12$}\,C_2(G)\right)\,R^b
\label{colourids}\eeq
where $R^a:=R(T^a)$ and $C_2(R)$ is the quadratic Casimir invariant of the
representation $R$ (with value $C_2(G)$ in the adjoint
representation). For $G=SU(N)$, one has $C_2(G)=N$ and
$C_2(N)=(N^2-1)/2N$ for the fundamental representation.

The cancellation of infrared divergences in loop amplitudes of QCD is
far more delicate than in QED, as there is no analog of the
Bloch-Nordsieck theorem in this case. The Kinoshita-Lee-Nauenberg
theorem guarantees that, at the end of any perturbative calculation,
these divergences must cancel for any appropriately defined physical
quantity. However, at a given order of perturbation theory, a physical
quantity typically involves both virtual and real emission
contributions that are separately infrared divergent. Already at
two-loop level these divergences have a highly intricate
structure. Their precise form is specified by the Catani colour-space
factorization formula which also provides an efficient way of
organising amplitudes into divergent parts, which ultimately drop out
of physical quantities, and finite contributions.

The computation of multi-gluon amplitudes in nonabelian gauge theory
is rather complicated when one uses polarization states of vector
bosons. A much more efficient representation of amplitudes is provided
by adopting a helicity (or circular polarization) basis for external
gluons. In the spinor-helicity formalism, one expresses positive and negative
helicity polarization vectors in terms of massless Weyl spinors
$|k^\pm\rangle:=\frac12\,(\id\pm\gamma_5)u_k=\frac12\,(\id\pm\gamma_5)v_k$
through
\beq
e_\mu^\pm(k;q)=\pm\,\frac{\bigl\langle
  q^\mp\bigl|\gamma_\mu\bigr|k^\mp\bigr
\rangle}{\sqrt2\,\bigl\langle q^\mp\bigm|k^\pm\bigr\rangle}
\label{helbasis}\eeq
where $q$ is an arbitary null reference momentum which drops out of
the final gauge invariant amplitudes. The spinor products are crossing
symmetric, antisymmetric in their arguments, and satisfy the
identities
\bea
\bigl\langle k_i^-\bigm| k_j^+\bigr\rangle\,\bigl\langle
k_j^+\bigm|k_i^-\bigr\rangle&=&2k_i\cdot k_j \ , \nn\\
\bigl\langle k_i^-\bigm| k_j^+\bigr\rangle\,
\bigl\langle k_l^-\bigm| k_r^+\bigr\rangle&=&
\bigl\langle k_i^-\bigm| k_r^+\bigr\rangle\,
\bigl\langle k_l^-\bigm| k_j^+\bigr\rangle+
\bigl\langle k_i^-\bigm| k_l^+\bigr\rangle
\bigl\langle k_j^-\bigm| k_r^+\bigr\rangle \ .
\label{helids}\eea
Any amplitude with massless external fermions and vector bosons can be
expressed in terms of spinor products. Conversely, the spinor products
offer the most compact representation of helicity amplitudes which can
be related to more conventional amplitudes described in terms of
Lorentz invariants. For loop amplitudes, one uses a dimensional regularization
scheme in which all helicity states are kept four-dimensional and only
internal loop momenta are continued to $D=4+\epsilon$ dimensions.

\bigskip

\sect{6. Computing Loop Integrals}
At the very heart of perturbative quantum field theory is the problem
of computing Feynman integrals for multi-loop scattering
amplitudes. The integrations typically involve serious technical
challenges and for the most part are intractable by straightforward
analytical means. We will now survey some of the computational
techniques that have been developed for calculating quantum
loop amplitudes which arise in the field theories considered
previously.

\smallskip

\subsect{Asymptotic expansion}
In many physical instances one is interested in scattering amplitudes
in certain kinematical limits. In this case one may perform an
asymptotic expansion of multi-loop diagrams whose coefficients are
typically non-analytic functions of the perturbative expansion parameter
$\hbar$. The main simplification which arises comes from the fact that
the expansions are done before any momentum integrals are
evaluated. In the limits of interest, Taylor series expansions in
different selected regions of each loop momentum can be interpreted in
terms of subgraphs and co-subgraphs of the original Feynman diagram.

Consider a Feynman diagram $\mathfrak{D}$ which depends on a
collection $\{Q_i\}$ of large momenta (or masses), and a collection
$\{m_i,q_i\}$ of small masses and momenta. The prescription for the
large momentum asymptotic expansion of $\mathfrak{D}$ may be
summarized in the diagrammatic formula
\beq
\lim_{Q\to\infty}\,\mathfrak{D}(Q;m,q)=\sum_{\mathfrak{d}\subset
\mathfrak{D}}\,(\mathfrak{D}/\mathfrak{d})(m,q)\star
\left(\mathcal{T}_{\{m_{\mathfrak{d}},q_{\mathfrak{d}}\}}\mathfrak{d}\right)
(Q;m_{\mathfrak{d}},q_{\mathfrak{d}}) \ ,
\label{asymptdiagformula}\eeq
where the sum runs through all subgraphs $\mathfrak{d}$ of
$\mathfrak{D}$ which contain all vertices where a large momentum
enters or leaves the graph and is one-particle irreducible after
identifying these vertices. The operator
$\mathcal{T}_{\{m_{\mathfrak{d}},q_{\mathfrak{d}}\}}$ performs a
Taylor series expansion before any integration is carried out, and the
notation $(\mathfrak{D}/\mathfrak{d})\star
(\mathcal{T}_{\{m_{\mathfrak{d}},q_{\mathfrak{d}}\}}\mathfrak{d})$
indicates that the subgraph $\mathfrak{d}\subset\mathfrak{D}$ is
replaced by its Taylor expansion in all masses and external momenta of
$\mathfrak{d}$ that do not belong to the set $\{Q_i\}$. The external
momenta of $\mathfrak{d}$ which become loop momenta in $\mathfrak{D}$
are also considered to be small. The loop integrations are then
performed only after all these expansions have been carried out. The
diagrams $\mathfrak{D}/\mathfrak{d}$ are called co-subgraphs.

The subgraphs become massless integrals in which the scales are set by the
large momenta. For instance, in the simplest case of a single large
momentum $Q$ one is left with integrals over propagators. The
co-subgraphs may contain small external momenta and masses, but the
resulting integrals are typically much simpler than the original
one. A similar formula is true for large mass expansions, with the
vertex conditions on subdiagrams replace by propagator conditions. For
example, consider the asymptotic expansion of the two-loop double
bubble diagram
\beq
\input{pert8.pstex_t}
\label{asymptex}\eeq
in the region $q^2\ll m^2$, where $m$ is the mass of the inner
loop. The subgraphs (to the right of the stars) are expanded in all
external momenta including $q$ and reinserted into the fat vertices of
the co-subgraphs (to the left of the stars). Once such asymptotic
expansions are carried out, one may attempt to reconstruct as much
information as possible about the given scattering amplitude by using the
method of Pad\'e approximation which requires knowledge of only part
of the expansion of the diagram. By construction, the Pad\'e
approximation has the same analytic properties as the exact
amplitude.

\smallskip

\subsect{Brown-Feynman reduction}
When considering loop diagrams which involve fermions or gauge bosons,
one encounters tensorial Feynman integrals. When these involve more
than three distinct denominator factors (propagators), they require
more than two Feynman parameters for their evaluation and become
increasingly complicated. The Brown-Feynman method simplifies such
higher-rank integrals and effectively reduces them to scalar integrals
which typically require fewer Feynman parameters for their evaluation.

To illustrate the idea behind this method, consider the one-loop
rank~$3$ tensor Feynman integral
\beq
J^{\mu\nu\lambda}=\int\frac{\dd^Dk}{(2\pi)^D}~
\frac{k^\mu\,k^\nu\,k^\lambda}{k^2\left(k^2-\mu^2\right)
(q-k)^2\bigl((k-q)^2+\mu^2\bigr)(k^2+2k\cdot p)}
\label{Jrank3def}\eeq
where $p$ and $q$ are external momenta with the mass-shell conditions
$p^2=(p-q)^2=m^2$. By Lorentz invariance, the general structure of the
integral (\ref{Jrank3def}) will be of the form
\beq
J^{\mu\nu\lambda}=a^{\mu\nu}\,p^\lambda+b^{\mu\nu}\,q^\lambda+
c^\mu\,s^{\nu\lambda}+c^\nu\,s^{\mu\lambda} \,
\label{Jgenform}\eeq
where $a^{\mu\nu}$, $b^{\mu\nu}$ are tensor-valued functions and
$c^\mu$ a vector-valued function of $p$ and $q$. The symmetric tensor
$s^{\mu\nu}$ is chosen to project out components of vectors transverse
to both $p$ and $q$, i.e. $p_\mu\,s^{\mu\nu}=q_\mu\,s^{\mu\nu}=0$,
with the normalization $s_\mu^{~\mu}=D-2$. Solving these constraints
leads to the explicit form
\beq
s^{\mu\nu}=\eta^{\mu\nu}-\frac{m^2\,q^\mu\,q^\nu+q^2\,p^\mu\,p^\nu-
(p\cdot q)\left(q^\mu\,p^\nu+p^\mu\,q^\nu\right)}{m^2\,q^2-(p\cdot q)^2} \ .
\label{symtensexpl}\eeq

To determine the as yet unknown functions $a^{\mu\nu}$, $b^{\mu\nu}$
and $c^\mu$ above, we first contract both sides of the decomposition
(\ref{Jgenform}) with $p^\mu$ and $q^\mu$ to get
\beq
2p_\lambda\,J^{\mu\nu\lambda}=2m^2\,a^{\mu\nu}+2(p\cdot q)\,b^{\mu\nu}
\ , ~~ 2q_\lambda\,J^{\mu\nu\lambda}=2(p\cdot q)\,a^{\mu\nu}+
2q^2\,b^{\mu\nu} \ .
\label{Jpqcontract}\eeq
Inside the integrand of (\ref{Jrank3def}), we then use the trivial
identities
\beq
2k\cdot p=\left(k^2+2k\cdot p\right)-k^2 \ , ~~
2q\cdot k=k^2+q^2-(k-q)^2
\label{trividspq}\eeq
to write the left-hand sides of (\ref{Jpqcontract}) as the sum of
rank~$2$ Feynman integrals which, with the exception of the one
multiplied by $q^2$ from (\ref{trividspq}), have one less denominator
factor. This formally determines the coefficients $a^{\mu\nu}$ and
$b^{\mu\nu}$ in terms of a set of rank~$2$ integrations. The vector
function $c^\mu$ is then found from the contraction
\beq
J^{\mu\nu}_{~~~\nu}=p_\nu\,a^{\mu\nu}+q_\nu\,b^{\mu\nu}+(D-2)\,c^\mu \ .
\label{Jcontra}\eeq
This contraction eliminates the $k^2$ denominator factor in the
integrand of (\ref{Jrank3def}) and produces a vector-valued
integral. Solving the system of algebraic equations
(\ref{Jpqcontract}) and (\ref{Jcontra}) then formally determines the
rank~$3$ Feynman integral (\ref{Jrank3def}) in terms of rank~$1$ and
rank~$2$ Feynman integrals. The rank~$2$ Feynman integrals thus
generated can then be evaluated in the same way by writing a
decomposition for them analogous to (\ref{Jgenform}) and solving for
them in terms of vector-valued and scalar-valued Feynman
integrals. Finally, the rank~$1$ integrations can be solved for in
terms of a set of scalar-valued integrals, most of which have fewer
denominator factors in their integrands.

Generally, any one-loop amplitude can be reduced to a set of basic
integrals by using the Passarino-Veltman reduction technique. For
example, in supersymmetric amplitudes of gluons any tensor Feynman
integral can be reduced to a set of scalar integrals, i.e. Feynman
integrals in a scalar field theory with a massless particle
circulating in the loop, with rational coefficients. In the case of
$\mathcal{N}=4$ supersymmetric Yang-Mills theory, only scalar box
integrals appear.

\smallskip

\subsect{Reduction to master integrals}
While the Brown-Feynman and Passarino-Veltman reductions are
well-suited for dealing with one-loop diagrams, they become rather cumbersome
for higher loop
computations. There are other more powerful methods for reducing
general tensor integrals into a basis of known integrals called master
integrals. Let us illustrate this technique on a scalar example. Any
scalar massless two-loop Feynman integral can be brought into the form
\beq
I(p)=\int\frac{\dd^Dk}{(2\pi)^D}~\int\frac{\dd^Dk'}{(2\pi)^D}~
\prod_{j=1}^t\,\Delta_j^{-l_j}~\prod_{i=1}^q\,\Sigma_i^{n_i} \ ,
\label{2loopgenform}\eeq
where $\Delta_j$ are massless scalar propagators depending on the loop
momenta $k,k'$ and the external momenta $p_1,\dots,p_n$, and
$\Sigma_i$ are scalar products of a loop momentum with an external
momentum or of the two loop momenta. The topology of the corresponding
Feynman diagram is uniquely determined by specifying the set
$\Delta_1,\dots,\Delta_t$ of $t$ distinct propagators in the graph,
while the integral itself is specified by the powers $l_j\geq1$ of all
propagators, by the selection $\Sigma_1,\dots,\Sigma_q$ of $q$ scalar
products and by their powers $n_i\geq0$.

The integrals in a class of diagrams of the same topology with the
same denominator dimension $r=\sum_jl_j$ and same total scalar product
number $s=\sum_in_i$ are related by various identities. One class
follows from the fact that the integral over a total
derivative with respect to any loop momentum vanishes in dimensional
regularization as $\int\frac{\dd^Dk}{(2\pi)^D}~\frac{\partial J(k)}{\partial
k^\mu}=0$, where $J(k)$ is any tensorial combination of propagators, scalar
products and loop momenta. The resulting relations are called
{\it integration by parts identities} and for two-loop integrals can be cast
into the form
\beq
\int\frac{\dd^Dk}{(2\pi)^D}~\int\frac{\dd^Dk'}{(2\pi)^D}~
v^\mu\,\frac{\partial f(k,k',p)}{\partial k^\mu}~=~0~=~
\int\frac{\dd^Dk}{(2\pi)^D}~\int\frac{\dd^Dk'}{(2\pi)^D}~
v^\mu\,\frac{\partial f(k,k',p)}{\partial k^{\prime\,\mu}} \ ,
\label{IBPids}\eeq
where $f(k,k',p)$ is a scalar function containing propagators and
scalar products, and $v^\mu$ is any internal or external momentum. For
a graph with $\ell$ loops and $n$ independent external momenta, this
results in a total of $\ell\,(n+\ell)$ relations.

In addition to these identities, one can also exploit the fact that all
Feynman integrals (\ref{2loopgenform}) are Lorentz scalars. Under an
infinitesimal Lorentz transformation $p^\mu\to p^\mu+\delta p^\mu$,
with $\delta p^\mu=p^\nu\,\delta\epsilon_\nu^{~\mu}$,
$\delta\epsilon_\nu^{~\mu}=-\delta\epsilon_\mu^{~\nu}$, one has the
invariance condition $I(p+\delta p)=I(p)$, which leads to the linear
homogeneous differential equations
\beq
\sum_{i=1}^n\left(p_i^\nu\,\frac\partial{\partial p_{i\,\mu}}-
p_i^\mu\,\frac\partial{\partial p_{i\,\nu}}\right)I(p)=0 \ .
\label{LIdiffeqs}\eeq
This equation can be contracted with all possible antisymmetric
combinations of $p_{i\,\mu}\,p_{j\,\nu}$ to yield linearly independent
{\it Lorentz invariance identities} for (\ref{2loopgenform}).

Using these two sets of identities, one can either obtain a reduction of
integrals of the type (\ref{2loopgenform}) to those corresponding to a
small number of simpler diagrams of the same topology and diagrams of
simpler topology (fewer denominator factors), or a complete
reduction to diagrams with simpler topology. The remaining integrals
of the topology under consideration are called {\it irreducible master
  integrals}. These momentum integrals cannot be further reduced and
have to be computed by different techniques. For instance, one can
apply a Mellin-Barnes transformation of all propagators given by
\beq
\frac1{(k^2+a)^l}=\frac1{(l-1)!}~\int\limits_{-\ii\infty}^{\ii\infty}
\frac{\dd z}{2\pi\ii}~\frac{a^z}{\left(k^2\right)^{l+z}}~\Gamma(l+z)\,
\Gamma(-z) \ ,
\label{MBtransf}\eeq
where the contour of integration is chosen to lie to the right of the poles
of the Euler function $\Gamma(l+z)$ and to the left of the poles of
$\Gamma(-z)$ in the complex $z$-plane.
Alternatively, one may apply the negative dimension
method in which $D$ is regarded as a negative integer in intermediate
calculations and the problem of loop integration is replaced with that
of handling infinite series. When combined with the above methods, it
may be used to derive powerful recursion relations among scattering
amplitudes. Both of these techniques rely on an
explicit integration over the loop momenta of the graph, their
differences occuring mainly in the representations used for the
propagators.

The procedure outlined above can also be used to reduce a tensor
Feynman integral to scalar integrals, as in the Brown-Feynman
and Passarino-Veltman reductions. The tensor integrals are expressed
as linear combinations of scalar integrals of either higher dimension
or with propagators raised to higher powers. The projection onto a
tensor basis takes the form (\ref{2loopgenform}) and can thus be
reduced to master integrals.

\bigskip

\sect{7. String Theory Methods}
The realizations of field theories as the low-energy limits of string
theory provides a number of powerful tools for the calculation of
multi-loop amplitudes. They may be used to provide sets of
diagrammatic computational rules, and they also work well for calculations
in quantum gravity. In this final part we shall briefly sketch the
insights into perturbative quantum field theory that are
provided by techniques borrowed from string theory.

\smallskip

\subsect{String theory representation}
String theory provides an efficient compact representation of scattering
amplitudes. At each loop order there is only a single closed string
diagram, which includes within it all Feynman graphs along with the
contributions of the infinite tower of massive string
excitations. Schematically, at one-loop order one has
\beq
\input{pert9.pstex_t}
\label{stringschem}\eeq
The terms arising from the heavy string modes are removed by taking
the low-energy limit in which all external momenta lie well below the
energy scale set by the string tension. This limit picks out the
regions of integration in the string diagram corresponding to
particle-like graphs, but with different diagrammatic rules.

Given these rules, one may formulate a purely field theoretic
framework which reproduces them. In the case of QCD, a key ingredient
is the use of a special gauge originally derived from the low-energy limit of
tree-level string amplitudes. This is known as the {\it Gervais-Neveu
gauge} and it is defined by the gauge-fixing lagrangian density
\beq
\mathcal{L}_{\rm GN}=-\mbox{$\frac12$}\,\Tr\left(\partial_\mu A^{\mu}
-\mbox{$\frac{\ii e}{\sqrt2}$}\,A_\mu\,A^\mu\right)^2 \ .
\label{GNgauge}\eeq
This gauge choice simplifies the colour factors that arise in
scattering amplitudes. The string theory origin of gauge theory
amplitudes is then most closely mimicked by combining this gauge with
the background field gauge, in which one decomposes the gauge
field into a classical background field and a fluctuating quantum
field as $A^{~}_\mu=A_\mu^{\rm cl}+A_\mu^{\rm qu}$, and imposes the
gauge-fixing condition $D_\mu^{\rm cl}A^{{\rm qu}\,\mu}=0$ where
$D_\mu^{\rm cl}$ is the background field covariant derivative
evaluated in the adjoint representation of the gauge group. This
hybrid gauge is well-suited for computing the effective action, with the
quantum part describing gluons propagating around loops and the
classical part describing gluons emerging from the loops. The leading
loop momentum behaviour of one-particle irreducible graphs with gluons
in the loops is very similar to that of graphs with scalar fields in
the loops.

\smallskip

\subsect{Supersymmetric decomposition}
String theory also suggests an intimate relationship with
supersymmetry. For example, at tree-level QCD is effectively
supersymmetric because a multi-gluon tree amplitude contains no fermion
loops, and so the fermions may be taken to lie in the adjoint
representation of the gauge group. Thus pure gluon tree-amplitudes in
QCD are identical to those in supersymmetric Yang-Mills theory. They
are connected by supersymmetric Ward identities to amplitudes with
fermions (gluinos) which drastically simplify computations. In
supersymmetric gauge theory these identities hold to all orders of
perturbation theory.

At one-loop order and beyond, QCD is not supersymmetric. However, one
can still perform a supersymmetric decomposition of a QCD amplitude
for which the supersymmetric components of the amplitude obey the
supersymmetric Ward identities. Consider, for example, a one-loop
multi-gluon scattering amplitude. The contribution from a fermion
propagating in the loop can be decomposed into the contribution of a
complex scalar field in the loop plus a contribution from an
$\mathcal{N}=1$ chiral supermultiplet consisting of a
complex scalar field and a Weyl fermion. The contribution from a gluon
circulating in the loop can be decomposed into contributions of a
complex scalar field, an $\mathcal{N}=1$ chiral supermultiplet, and an
$\mathcal{N}=4$ vector supermultiplet comprising three complex scalar
fields, four Weyl fermions and one gluon all in the adjoint
representation of the gauge group. This decomposition assumes the use
of a supersymmetry preserving regularization.

The supersymmetric components have important cancellations in their
leading loop momentum behaviour. For instance, the leading large loop
momentum power in an $n$-point 1PI graph is reduced from $|k|^n$ down
to $|k|^{n-2}$ in the $\mathcal{N}=1$ amplitude. Such a reduction can
be extended to any amplitude in supersymmetric gauge theory and is
related to the improved ultraviolet behaviour of supersymmetric
amplitudes. For the $\mathcal{N}=4$ amplitude, further cancellations
reduce the leading power behaviour all the way down to $|k|^{n-4}$. In
dimensional regularization, $\mathcal{N}=4$ supersymmetric loop amplitudes
have a very simple analytic structure owing to their origins as the
low-energy limits of superstring scattering amplitudes. The
supersymmetric Ward identities in this way can be used to provide
identities among the non-supersymmetric contributions. For example, in
$\mathcal{N}=1$ supersymmetric Yang-Mills theory one can deduce that
fermion and gluon loop contributions are equal and opposite for
multi-gluon amplitudes with maximal helicity violation.

\smallskip

\subsect{Scattering amplitudes in twistor space}
The scattering amplitude in QCD with $n$ incoming gluons of the same
helicity vanishes, as does the amplitude with $n-1$ incoming gluons of
one helicity and one gluon of the opposite helicity for $n\geq3$. The first
non-vanishing amplitudes are the {\it maximal helicity violating
(MHV)} amplitudes involving $n-2$ gluons of one helicity and two
gluons of the opposite helicity. Stripped of the momentum conservation
delta-function and the group theory factor, the tree-level amplitude
for a pair of gluons of negative helicity is given by
\beq
\mathcal{A}(k)=e^{n-2}\,\left\langle\left. k_r^-\right|k_s^+\right
\rangle\,\prod_{i=1}^n\left\langle\left. k_i^-\right|k_{i+1}^+\right
\rangle^{-1} \ .
\label{gluonMHVampl}\eeq
This amplitude depends only on the holomorphic (negative chirality)
Weyl spinors. The full MHV amplitude (with the momentum conservation
delta-function) is invariant under the conformal group $SO(4,2)\cong
SU(2,2)$ of four dimensional Minkowski space. After a Fourier
transformation of the positive chirality components, the complexification
$SL(4,\complex)$ has an obvious four dimensional representation acting
on the positive and negative chirality spinor products. This
representation space is isomorphic to $\complex^4$ and is called {\it
twistor space}. Its elements are called {\it twistors}.

Wavefunctions and amplitudes have a known behaviour under the
$\complex^\times$-action which rescales twistors, giving the
projective twistor space $\complex\mathbb{P}^3$ or $\real\mathbb{P}^3$
according to whether the twistors are complex-valued or
real-valued. The Fourier transformation to twistor
space yields (due to momentum conservation) the localization of an MHV
amplitude to a genus~$0$ holomorphic curve
$\complex\mathbb{P}^1$ of degree~$1$ in $\complex\mathbb{P}^3$ (or to
a real line $\real\mathbb{P}^1\subset\real\mathbb{P}^3$). It is
conjectured that, generally, an $\ell$-loop amplitude with $p$ gluons of
positive helicity and $q$ gluons of negative helicity is supported on
a holomorphic curve in twistor space of degree $q+\ell-1$ and genus $\leq
\ell$. The natural interpretation of this curve is as the worldsheet of a
string. The perturbative gauge theory may then be described in terms
of amplitudes arising from the couplings of gluons to a string. This
twistor string theory is a topological string theory which gives the
appropriate framework for understanding the twistor properties of
scattering amplitudes. This framework has been used to analyse MHV
tree diagrams and one-loop $\mathcal{N}=4$ supersymmetric amplitudes
of gluons.

\bigskip

\sect{See Also}
Effective field theories. Perturbative renormalization theory and
BRST. Analytic properties of the S-matrix. Dispersion
relations. Scattering: Asymptotic completeness and bound
states. Scattering: Fundamental concepts and tools. Stationary phase
approximation. Supersymmetric particle models. Gauge theories from
strings.

\bigskip

\noindent
{\bf Further Reading}

\noindent
Bern, Z., Dixon, L. and Kosower, D.A. (1996), Progress in one-loop
QCD computations, {\it Annual Review of Nuclear and Particle Science} 46:
109--148.

\noindent
Brown, L.M. and Feynman, R.P. (1952), Radiative corrections to Compton
scattering, {\it Physical Review} 85: 231--244.

\noindent
Cachazo, F. and Svr\v{c}ek, P. (2005), Lectures on twistor strings and
perturbative Yang-Mills theory, Preprint arXiv:hep-th/0504194.

\noindent
Catani, S. (1998), The singular behaviour of QCD amplitudes at
two-loop order, {\it Physics Letters} B427: 161--171.

\noindent
Chetyrkin, K.G. and Tkachov, F.V. (1981), Integration by parts: The
algorithm to calculate beta functions in 4 loops, {\it Nuclear
Physics} B192: 159--204.

\noindent
Dixon, L. (1996), Calculating scattering amplitudes efficiently,
{\it QCD and Beyond}, World Scientific: 539--584.

\noindent
Fleischer, J., Jegerlehner, F. and Tarasov, O.V. (2000), Algebraic
reduction of one-loop Feynman graph amplitudes, {\it Nuclear Physics}
B566: 423--440.

\noindent
Gehrmann, T. and Remiddi, E. (2000), Differential equations for
two-loop four-point functions, {\it Nuclear Physics} B580: 485--518.

\noindent
Halliday, I.G. and Ricotta, R.M. (1987), Negative dimensional
integrals 1: Feynman graphs, {\it Physics Letters} B193: 241--246.

\noindent
Itzykson, C. and Zuber, J.-B. (1980), {\it Quantum Field Theory},
McGraw-Hill.

\noindent
Mangano, M.L. and Parke, S.J. (1991), Multiparton amplitudes in gauge
theories, {\it Physics Reports} 200: 301--367.

\noindent
Passarino, G. and Veltman, M.J.G. (1979), One-loop corrections for
$e^+e^-$ annihilation into $\mu^+\mu^-$ in the Weinberg model, {\it
Nuclear Physics} B160: 151--207.

\noindent
Peskin, M.E. and Schroeder, D.V. (1995), {\it An Introduction to
  Quantum Field Theory}, Addison-Wesley.

\noindent
Pokorski, S. (1987), {\it Gauge Field Theories}, Cambridge University
Press.

\noindent
Steinhauser, M. (2002), Results and techniques of multiloop
calculations, {\it Physics Reports} 364: 247--357.

\noindent
Weinberg, S. (1995), {\it The Quantum Theory of Fields, Volume 1:
Foundations}, Cambridge University Press.

\bigskip

\sect{Acknowledgments}
This work was supported in part by a PPARC Advanced Fellowship, by
PPARC Grant PPA/G/S/2002/00478, and by the EU-RTN Network Grant
MRTN-CT-2004-005104.

\end{document}